\documentclass[aps,prl,reprint,showpacs]{revtex4-1}
\usepackage{graphicx}
\usepackage{amssymb,amsmath}
\usepackage{dcolumn}

\usepackage{color}
\begin{document}

\title{Dimensional-Crossover-Driven Mott Transition in the Frustrated Hubbard Model}
    
\author{Marcin Raczkowski}
\author{Fakher F. Assaad}
\affiliation{Institut f\"ur Theoretische Physik und Astrophysik,
             Universit\"at W\"urzburg, Am Hubland, D-97074 W\"urzburg, Germany}

\date{\today}

\begin{abstract}
We study the Mott transition in a frustrated Hubbard model with next-nearest neighbor hopping at half-filling. 
The interplay between interaction, dimensionality and geometric frustration closes the one-dimensional Mott gap
and gives rise to a  metallic  phase with  Fermi surface pockets.  We argue that they emerge as a consequence 
of remnant one-dimensional Umklapp  scattering at the momenta  with  vanishing 
interchain hopping matrix elements. 
In this pseudogap  phase, enhanced $d$-wave pairing correlations are driven by antiferromagnetic 
fluctuations. Within the adopted cluster dynamical mean-field theory on the $8\times 2$ cluster and down 
to our lowest temperatures the transition from one to two dimensions is continuous. 
\end{abstract}

\maketitle

The relative importance of spatial versus local fluctuations in the understanding of the Mott transition~\cite{Imada98}   
can be tuned with dimensionality.  Starting from the high dimensional limit, 
experimental studies on V$_2$O$_3$ indicate that exactly as a usual gas-liquid transition, 
the three dimensional bandwidth-controlled Mott transition belongs to the conventional Ising 
universality class: it is a first-order transition below the  critical endpoint at $T_c\simeq 450$ K 
and affects solely the charge sector~\cite{Lim03}.  In contrast, Ising universality class has been 
ruled out in two-dimensional (2D) organic salts 
$\kappa$-(BEDT-TTF)$_2$X (BEDT-TTF: bis(ethylenedithio)tetrathiafulvalene, X: monovalent anion) 
with a much lower critical point $T_c\simeq 40$ K~\cite{Kag05,Kag09}. 
In this case geometric frustration, inherent to the triangular lattice, strengthens 
spin fluctuations which in turn affect the nature of the transition. 
The unconventional character of the quantum criticality in $\kappa$-(BEDT-TTF)$_2$X has been 
confirmed in recent numerical simulations~\cite{Sent11,Sem11,Sato11}.     
Moreover, enhanced spin fluctuations and spatial correlations in the copper oxide planes offer a natural framework 
which accounts for a depletion of low-energy states in the pseudogap regime of the high-$T_c$ 
superconductors~\cite{Sene04,Civ05,Macr06}. 
As for the one-dimensional (1D) regime, it is dominated by spatial fluctuations~\cite{Giamarchi_book}.  
The relevance of Umklapp scattering for the half-filled band  leads to the absence of a bandwidth-controlled  
Mott transition. However, a Mott transition can be triggered as a function of dimensionality.   

The aim of this Letter is to reexamine the dimensional-crossover-driven Mott
transition in the quasi-1D Hubbard model at half-filling.
The subject combines many fascinating  issues such as the breakdown of 
spin-charge separation and the binding of spinons into magnons~\cite{Lake10}. These phenomena follow 
from a delicate interplay between  $\pmb k$-space and temporal  fluctuations.     
To capture the relevant physics we employ a cluster extension of the dynamical mean-field 
theory (CDMFT)~\cite{Geo96}. In the CDMFT, a cluster of $N_c$ impurities is subject to a dynamical effective 
bath simulating the effect of all the other sites of the lattice~\cite{Kot01}. The ability of the CDMFT to reproduce 
the density-driven Mott transition in the 1D Hubbard model has been demonstrated 
in Refs.~\cite{Bol03,Cap04}. Previous CDMFT studies of the dimensional-crossover-driven Mott transition 
in weakly-coupled 1D \emph{chains} yielded ordinary open Fermi surface (FS) in the Hubbard model~\cite{Bier01} 
and small FS pockets in the model of \emph{spinless} fermions~\cite{Bert06}.

\begin{figure}[t!]
\begin{center}
\unitlength=0.01\textwidth
\begin{picture}(45,27)
\put(0,0){\includegraphics[width=0.45\textwidth]{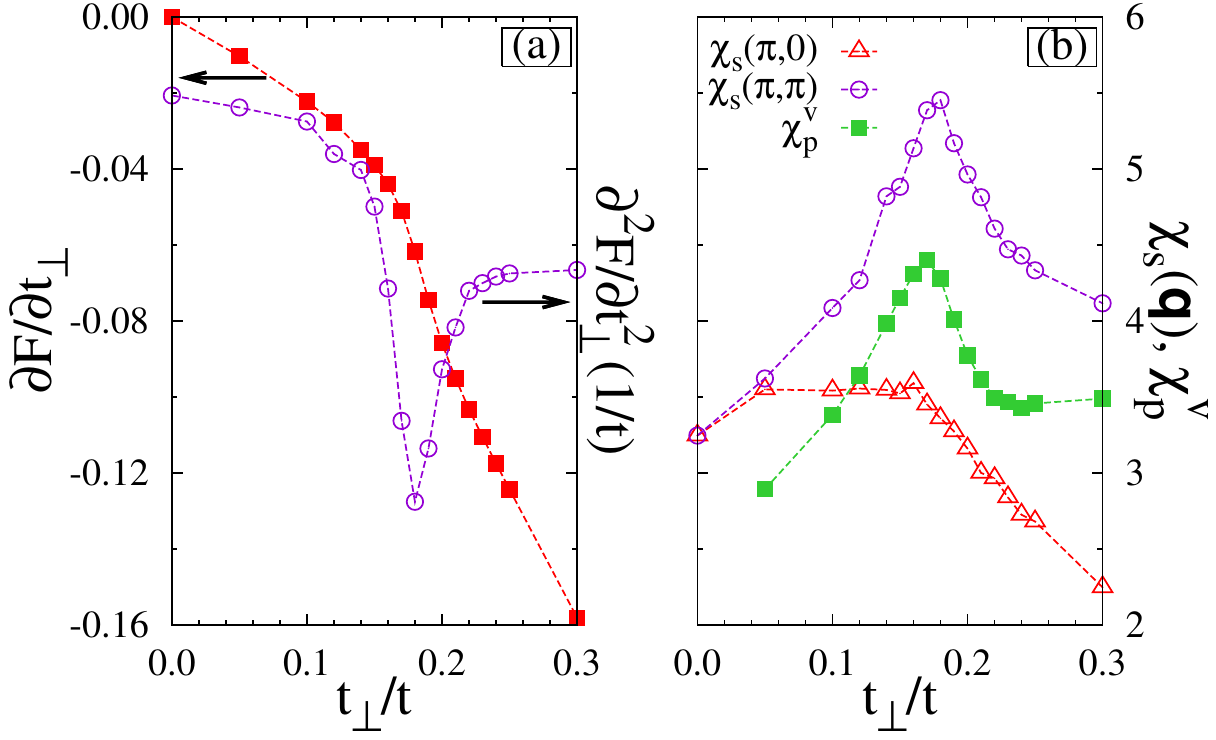}}
\put(7.,5.){\includegraphics[width=0.075\textwidth]{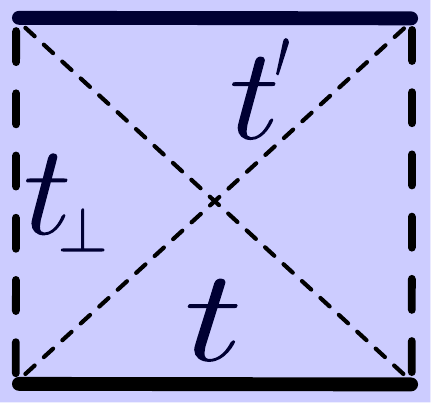}}
\end{picture}
\end{center}
\caption {(color online)
(a) $\partial F/\partial t_{\perp}$ (squares) and susceptibility of interchain charge fluctuations
$\partial^2 F/\partial t_{\perp}^2$ (scaled by a factor 0.1, circles);
(b) spin susceptibility $\chi_s({\pmb q})$, and interaction vertex contribution to the $d$-wave
pairing susceptibility $\chi_p^v$ (multiplied by 100) on the $8\times 2$ cluster. Inset shows the lattice
geometry of the Hamiltonian (\ref{eq:Hubb}). Parameters: $U/t=3$, $t'=-t_{\perp}/4$ and $T=t/30$.}
\label{fig:mott}
\end{figure}

We study the Hubbard model on a strongly anisotropic square lattice at half-filling, 
\begin{equation}
H=-\sum_{\pmb{ij},\sigma}t^{}_{\pmb{ij}}
   c^{\dag}_{{\pmb i}\sigma}c^{}_{{\pmb j}\sigma} +
   U\sum_{\pmb i}n^{}_{{\pmb i}\uparrow}n^{}_{{\pmb i}\downarrow} 
   -\mu\sum_{\pmb i,\sigma}n_{{\pmb i}\sigma},
\label{eq:Hubb}
\end{equation}
where the electron hopping $t_{\pmb{ij}}$ is $t$ ($t_{\perp}$) on the intrachain (interchain) 
bonds, $\mu$ is the chemical potential and we set $U/t=3$. In addition, we
allow for a finite diagonal next-nearest neighbor hopping 
$t'=-t_{\perp}/4$ (see Fig.~\ref{fig:mott}). 
It brings about \emph{frustration} in the ground state and by reducing nesting properties
of the FS precludes long-range magnetic order in the $T=0$ and weak-coupling regime~\cite{Tsu07, Mou11}.
Hence, finite value of $t'$ guarantees the Mott transition in the thermodynamic limit.
We use the Hirsch-Fye Quantum Monte Carlo (QMC) algorithm as a
cluster-impurity solver and extend previous studies~\cite{Bier01} to a low temperature regime. 
However, computational cost prevented us from decreasing the temperature below $T =t/30$ 
on the $8\times 2$ cluster~\cite{sign}. The CDMFT allows one to compute the single-particle spectral function
$A({\pmb k},\omega) = - \tfrac{1}{\pi} {\rm Im} G({\pmb k}, \omega )$. Here $G({\pmb k}, \omega )$ is
the lattice Green's function represented in the original Brillouin zone. We estimate the latter
by periodizing the cluster Green's function and applying stochastic analytical continuation of the QMC
data~\cite{Beach04a}.

\begin{figure}[t!]
\begin{center}
\includegraphics[width=0.45\textwidth]{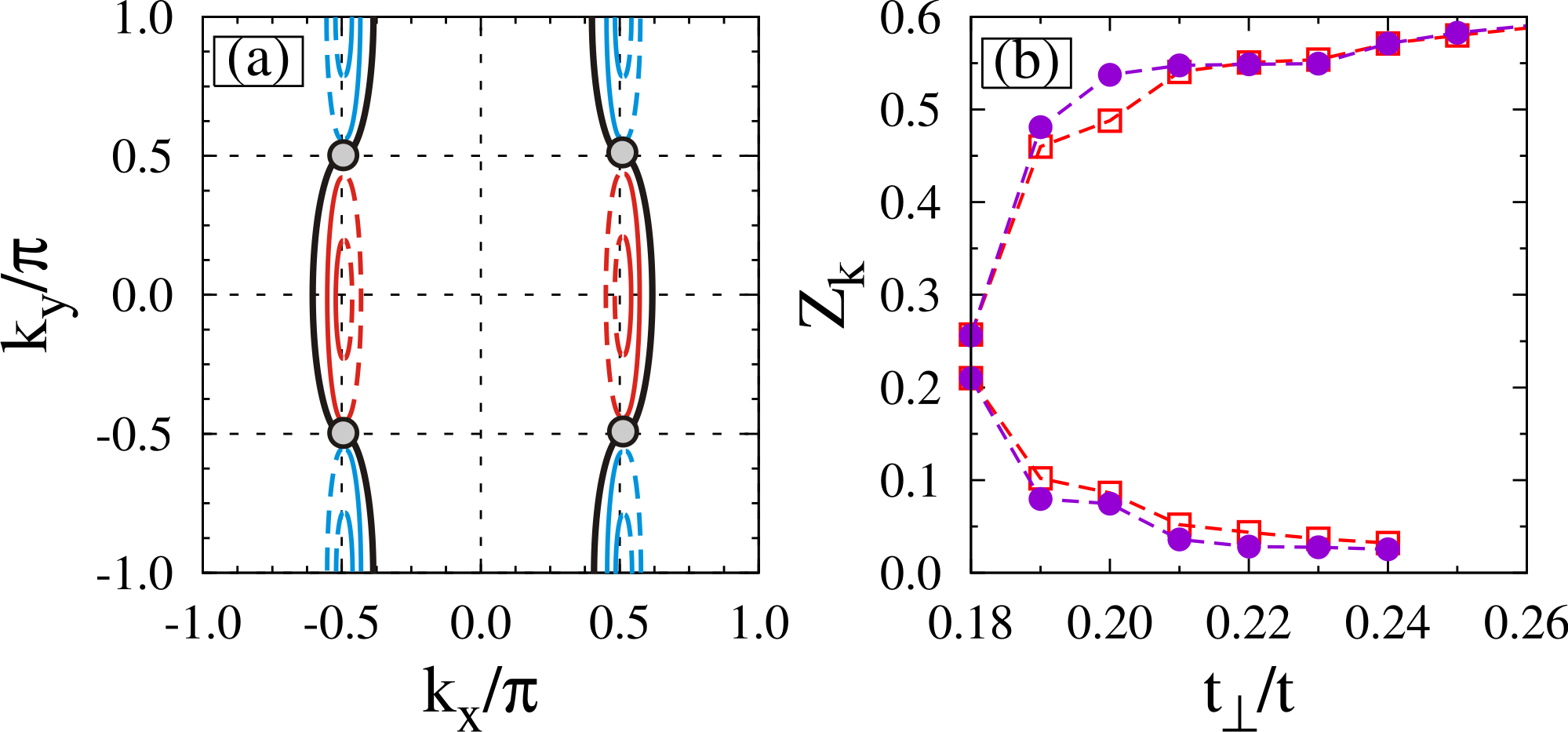}
\end{center}
\caption {(color online)
(a) Schematic evolution of the FS. 
The emergent FS forms \emph{electron} pockets around ${\pmb k}=(\pm\pi/2,0)$
and \emph{hole} pockets around ${\pmb k}=(\pm\pi/2,\pm\pi)$.
The pockets have anisotropic distribution of the spectral weight
between the \emph{main} (solid) and \emph{ghost} (dashed) vertical segments. The anisotropy and
pockets' size grow until $t_{\perp}/t$=0.25 when the ghost segments
vanish and the main ones merge into the open FS of a quasi-1D metal (bold).
The QP weight retains a strong ${\pmb k}$-dependence with a minimum
at hot spots (dots).
(b) QP weight $Z_{{\pmb k}}$ on the main and ghost sides of the electron 
(squares) and hole (circles) pockets.}
\label{fig:FS}
\end{figure}

Our main results are summarized in Figs.~\ref{fig:mott} and \ref{fig:FS}.
The control parameter $t_{\perp}$ interpolates between one and two dimensions and triggers the  Mott transition at 
$t_{\perp}/t\simeq 0.18$. To pin down the nature of the transition, continuous or first order,  we  plot in 
Fig.~\ref{fig:mott}(a) 
$\tfrac{\partial F}{\partial t_{\perp}}    = \tfrac{1}{N}\sum_{{\pmb k}\sigma}
\tfrac{\partial\varepsilon_{\pmb k}}{\partial t_{\perp}}\langle 
c_{{\pmb k}\sigma}^{\dagger} c_{{\pmb k}\sigma}^{}\rangle$ with 
$\varepsilon_{\pmb k}= -2(t\cos k_x +t_{\perp}\cos k_y) + t_{\perp} \cos k_x\cos k_y$. 
Down to the considered  temperatures,  we do not detect  a   jump and thus conclude that the transition is continuous.
$\partial^2 F/\partial t_{\perp}^2$  in Fig.~\ref{fig:mott}(a)   corresponds to  an interchain charge 
susceptibility  which is  greatly enhanced in the vicinity of the  critical coupling $t_{\perp}/t=0.18$. 
The open warped Fermi lines which form at sufficiently large $t_{\perp}/t>0.24$ (see Fig.~\ref{fig:FS}) essentially follow 
from the topology of the tight-binding model. In the intermediate region, $
0.18\leqslant t_{\perp}/t\leqslant 0.24 $,  
we find a metallic phase where the FS is broken into \emph{electron} and \emph{hole} pockets.  
On the one hand, starting from the 1D Mott insulating state, the occurrence  of the  pockets might be understood  by 
taking the interchain hopping into account at the random-phase approximation (RPA) level~\cite{Ess02}. 
In this context the \emph{nodal} points ${\pmb k}=(\pm\pi/2,\pm\pi/2)$ play a special role since  
$\frac{\partial \varepsilon_{\pmb k} } {\partial t_{\perp}  } = 0$ there.    
On the other hand, starting from the large $t_{\perp}$ limit, scattering off   
 ${\pmb q}_1=(\pi,0)$ and/or $\pmb{q}_2 = (\pi,\pi)$ magnetic fluctuations could equally gap out the hot spots,  
${\pmb k}=(\pm\pi/2,\pm\pi/2)$.  
To provide support for this scenario, we plot in Fig.~\ref{fig:mott}(b) the cluster spin susceptibility
$\chi_s({\pmb q})=
\tfrac{1}{N_c}\int_{0}^{\beta}d\tau\sum_{\pmb{ij}} e^{i{\pmb q}({\pmb i}-{\pmb j})}
\langle\pmb{ S_i}(\tau)\pmb{S_j}(0)\rangle$ for both momenta. As apparent, 1D fluctuations ${\pmb q}_1=(\pi,0)$  
remain robust up to   $t_{\perp}/t = 0.16$ but are then gradually suppressed, giving way to dominant  $\pmb{q}_2 = (\pi,\pi)$  
fluctuations, which peak at the Mott transition. Let us however note that in the static mean-field limit, 
antiferromagnetic order is not sufficient to reproduce the observed  FS topology pointing towards the remnant 
1D Umklapp  scattering at the nodal momenta as the origin of the pockets.

\begin{figure}[b!]
\begin{center}
\includegraphics[width=0.45\textwidth]{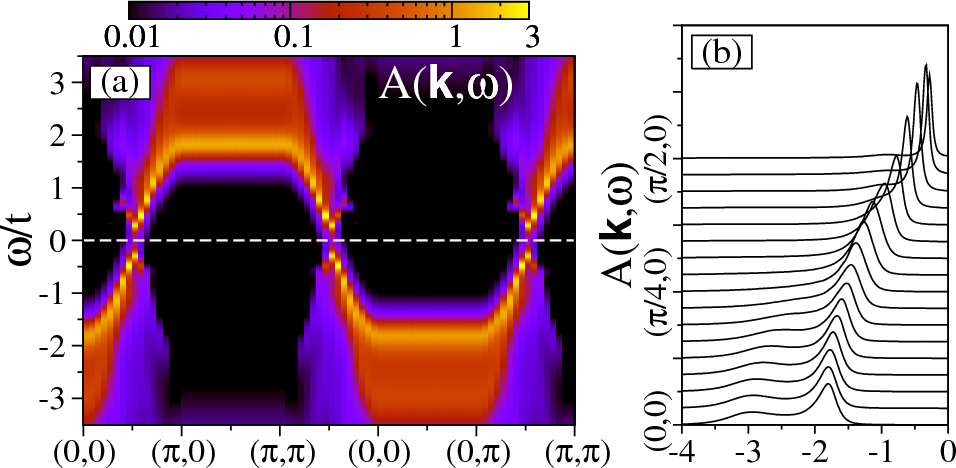}\\
\includegraphics[width=0.45\textwidth]{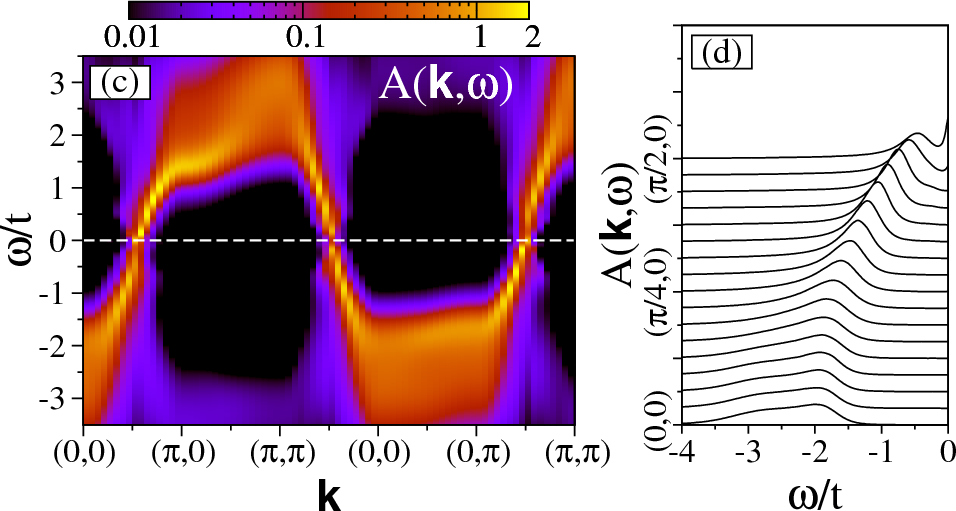}\\
\end{center}
\caption {(color online)
Dimensional-crossover-driven Mott transition as seen in the single-particle spectral function $A({\pmb k},\omega)$: 
(a,b) $t_{\perp}=0$ and (c,d) $t_{\perp}/t=0.18$. In panel (b), the spinon (holon) branch corresponds to the lower 
(higher) binding-energy peak, respectively. }
\label{fig:Ak}
\end{figure}

Enhanced staggered magnetic fluctuations $\propto\chi_s({\pi,\pi})$  give rise to pairing with a $d$-wave character
$\Delta^{\dagger}_{\pmb i}(\tau)=\pm\tfrac{1}{\sqrt{2}}[
c_{{\pmb i}\uparrow}^{\dagger}(\tau) c_{{\pmb i}+{\pmb\delta} \downarrow}^{\dagger}(\tau) -
c_{{\pmb i}\downarrow}^{\dagger}(\tau) c_{{\pmb i}+{\pmb\delta} \uparrow}^{\dagger}(\tau)]$
with the upper (lower) sign corresponding to ${\pmb\delta}={\pmb a}_x$ (${\pmb a}_y$),
respectively.
The response of the system in the particle-particle channel is best seen in the
pairing interaction vertex $\chi_p^v$~\cite{White89}. It is obtained from the full pairing
susceptibility
$\chi_p=
\tfrac{1}{N_c}\int_{0}^{\beta}d\tau\sum_{\pmb{ij}}
\langle\Delta^{\dagger}_{\pmb i}(\tau)\Delta_{\pmb j}(0)\rangle$
by subtracting  the uncorrelated contribution.
As shown in Fig.~\ref{fig:mott}(b), the calculated pairing
vertex $\chi_p^v$ is enhanced at the Mott transition which confirms the magnetic-pairing scenario.

\begin{figure}[t!]
\begin{center}
\includegraphics[width=0.45\textwidth]{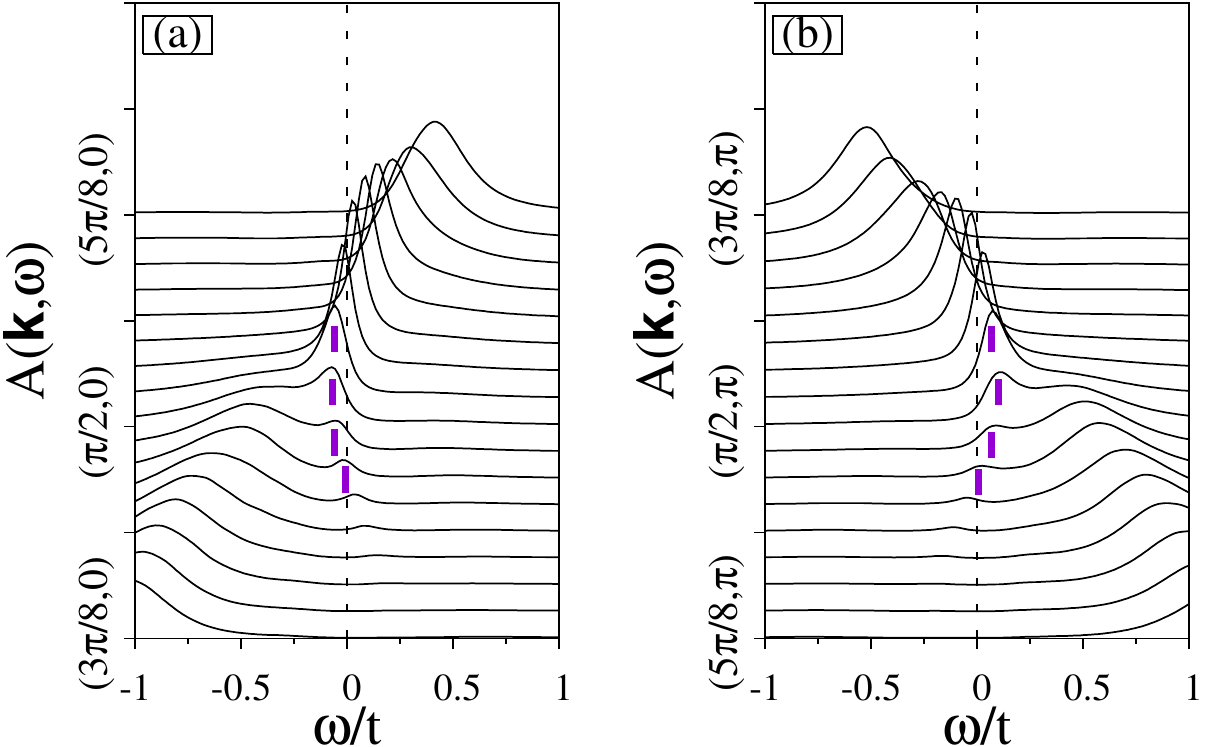}
\end{center}
\caption {(color online)
Low-energy part of the spectral function $A({\pmb k},\omega)$ at $t_{\perp}/t=0.2$.
Vertical bars track the position of the peaks constituting:
(a) electron and (b) hole FS pockets.}
\label{fig:pockets}
\end{figure}

The schematic evolution of the FS surface shown in Fig.~\ref{fig:FS}(a)  stems from the calculation of the 
single-particle Green's function. In Fig.~\ref{fig:Ak} we show the dimensional-crossover-driven Mott transition 
as seen in the evolution of the single-particle spectral function $A({\pmb k},\omega)$. 
Most noteworthy features in the 1D limit shown in Fig.~\ref{fig:Ak}(a,b)  are: 
(i) a well defined  single-particle gap at ${\pmb k}=(\pi/2,0)$ and  equivalent points;
(ii) signatures of  \emph{spinon} and \emph{holon} branches especially in the vicinity of  
${\pmb k}=(0,0)$~\cite{Mat05}, and 
(iii) backfolding of the energy bands around ${\pmb k}=(\pi/2,0)$ and the equivalent points.
Concerning (ii), already when $t_{\perp}/t=0.15$ the intensity of the spinon excitation is noticeably 
reduced but nevertheless one can distinguish two peaks~\cite{epaps1}. As shown in Fig.~\ref{fig:Ak}(c,d), they are 
superseded by a single peak with a broad shoulder at the Mott transition.   
Finally, a broad quasiparticle (QP) peak is resolved at $t_{\perp}/t=0.2$~\cite{epaps1}.
This remnant aspect of the 1D physics is captured in approaches starting from a fractionalized spectral 
function in the 1D limit and treating the interchain hopping  at the RPA level \cite{Ess02,Rib11}.
As for (iii), the pocket emerges when one of the backfolded bands intersects the Fermi energy. 
This defines a \emph{main} and \emph{ghost} side of the pocket which we can characterize with the magnitude 
of the QP residue $Z_{\pmb k}$. We extract this quantity  by fitting the  data to the Lorentzian form
$A({\pmb k},\omega)\simeq \tfrac{1}{\pi}
\tfrac{Z_{\pmb k}\Gamma_{\pmb k}}{(\omega-\varepsilon_{\pmb k}+\mu)^2+{\Gamma}_{\pmb k}^2}$ and plot it in Fig.~\ref{fig:FS}(b).
At the exception of Mott transition at $t_{\perp}/t=0.18$ where the pockets shrink and become very thin,  strong anisotropy 
along the pockets is evident. We illustrate this in Fig.~\ref{fig:pockets} by 
showing the low-frequency part of $A({\pmb k},\omega)$ across the pockets at $t_{\perp}/t=0.2$.  
The two features --- one with a large and the second one with a small QP weight crossing  the Fermi level are part 
of the electron [Fig.~\ref{fig:pockets}(a)] and hole
[Fig.~\ref{fig:pockets}(b)] pocket. The broken FS is unrelated to a specific
ladder geometry of the $8\times 2$ cluster and it is also found on the square $4\times 4$ cluster~\cite{epaps2}.

\begin{figure}[t!]
\begin{center}
\includegraphics[width=0.45\textwidth]{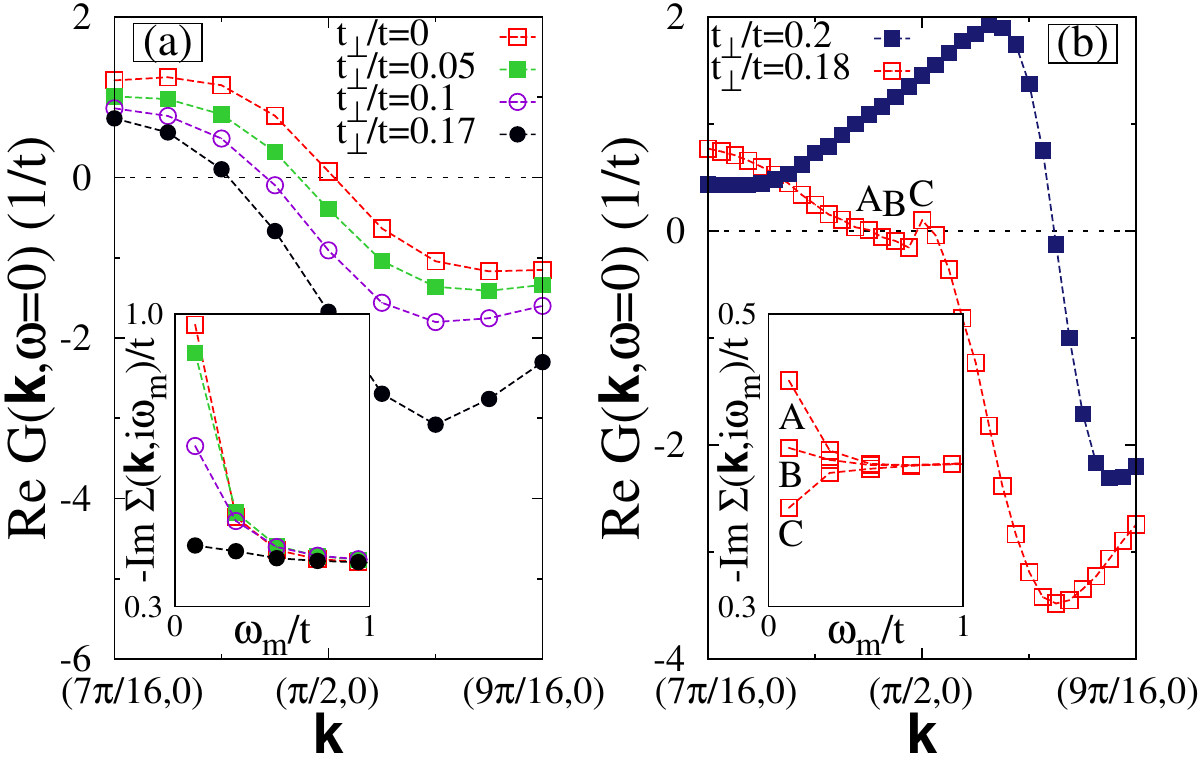}
\end{center}
\caption {(color online)
Real part of the zero-frequency Green's function around momentum ${\pmb k}=(\pi/2,0)$ in the :
(a) insulating Mott phase and (b) metallic phase. Insets show the low-frequency imaginary
part of the self-energy at: ${\pmb k}=(\pi/2,0)$ in panel (a)
and ${\pmb k}$ indicated by the capital letters in panel (b). $G({\pmb k},0)$ for $t_{\perp}/t=0.2$
was rescaled by 0.5.}
\label{fig:GA}
\end{figure}

The  reconstruction of the FS is governed  by the topology of the real part of the zero-frequency Green's function
$G({\pmb k},0)$~\cite{Stan06,Sak09,Gros12}. In the Fermi liquid theory, $G({\pmb k},0)$ is 
positive (negative) inside (outside) the FS, respectively, and changes sign by going through 
a pole. This contrasts with the Mott insulator in which $G({\pmb k},0)$ changes sign
in momentum space passing through a zero as a result of a diverging lattice self-energy
$\Sigma({\pmb k},i\omega_m)$. We extract the latter from the Dyson's equation
$G^{-1}({\pmb k}, i \omega_m) = G_0^{-1}({\pmb k}, i \omega_m)-\Sigma({\pmb k},i \omega_m)$,
where $G_0({\pmb k}, i \omega_m)$ is the bare Green's function, in combination with the
spectral representation of the lattice Green's function
       $G({\pmb k}, i \omega_m) =
        \int {\rm d} \omega \tfrac{ A({\pmb k},\omega) } { i \omega_m - \omega }$. 
The diverging behavior of $\Sigma({\pmb k},i\omega_m)$ at momentum ${\pmb k}=(\pi/2,0)$ in the Mott phase is 
shown in Fig.~\ref{fig:GA}(a).
It is the coexistence of infinities and zeros which accounts for the anisotropy of the emergent pockets. 
We focus on the electron pocket and illustrate this in Fig.~\ref{fig:GA}(b) for $t_{\perp}/t=0.18$. 
On the one hand, in close vicinity of the $C$-point with a vanishing $\Sigma({\pmb k},i\omega_m\to 0)$,  
two adjacent poles of $G({\pmb k},0)$ result in a very thin electron pocket. On the other hand, diverging 
$\Sigma({\pmb k},i\omega_m\to 0)$ yields a zero of $G({\pmb k},0)$ at the $A$-point.
The latter reduces the QP weight of the nearby ghost side.
The interference of the neighboring pole and zero becomes stronger with growing $t_{\perp}$ 
and prevented us from resolving the full structure of $G({\pmb k},0)$ already at $t_{\perp}/t=0.2$. Indeed, 
at our lowest temperature $T=t/30$ only a pole associated with the main side of the pocket and
a broad minimum in $G({\pmb k},0)$ is observed in Fig.~\ref{fig:GA}(b). However, as depicted in 
Fig.~\ref{fig:pockets}(a) the ghost side remains visible in the spectral function. 
In analogy with the density-driven Mott transition in the 2D Hubbard model~\cite{Stan06,Sak09}, we believe that 
the emergence of a large FS above $t_{\perp}/t=0.24$ corresponds to a simultaneous annihilation of 
the adjacent zero and pole leaving the pole carrying a larger QP weight.

\begin{figure}[t!]
\begin{center}
\includegraphics[width=0.45\textwidth]{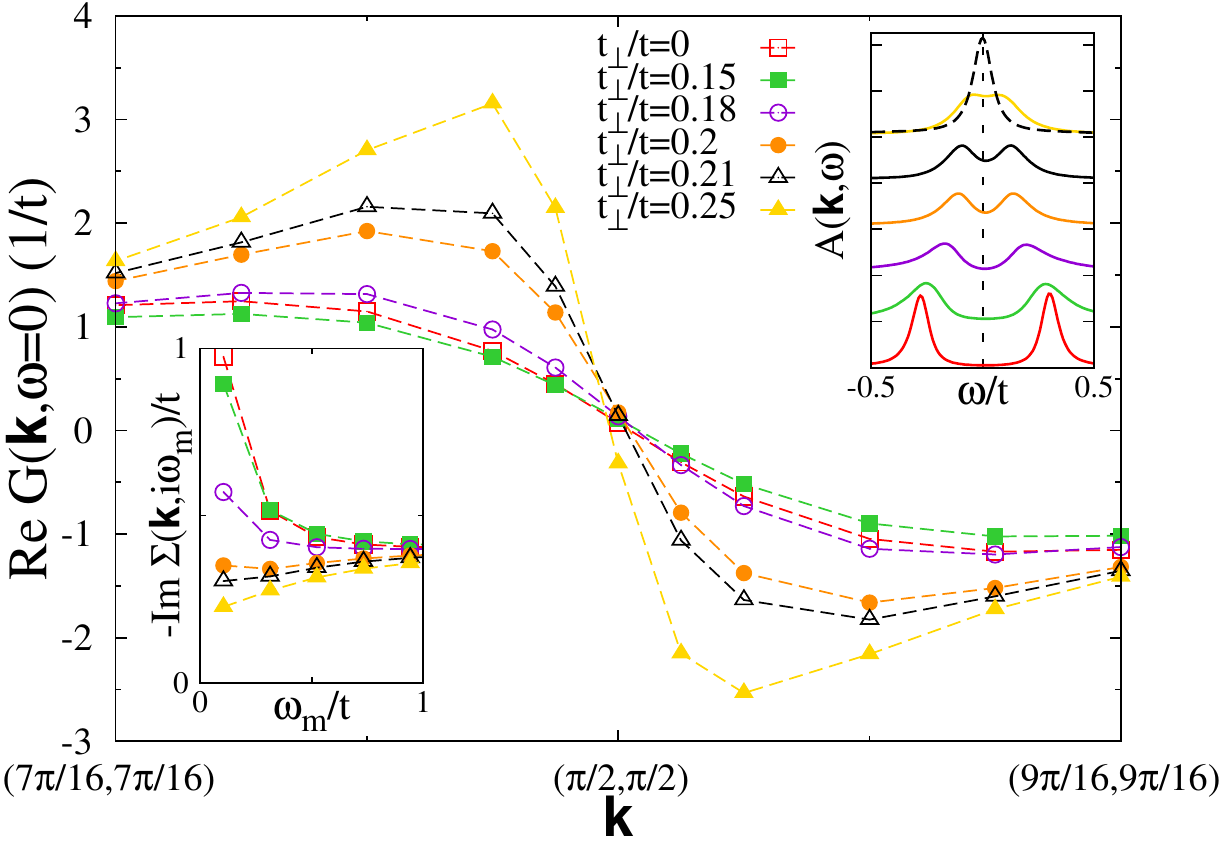}
\end{center}
\caption {(color online)
Real part of the zero-frequency Green's function near the nodal ${\pmb k}=(\pi/2,\pi/2)$ point. 
Insets show the low-frequency dependence of the imaginary part of the self-energy at ${\pmb k}=(\pi/2,\pi/2)$ 
(left) and the corresponding  spectral function from bottom ($t_{\perp} =0) $ to top ($t_{\perp}/t =0.25$) (right).
For comparison, dashed line in the right inset shows $A({\pmb k},\omega)$ at ${\pmb k}_F$ along the 
$(0,0)-(\pi,0)$ direction at $t_{\perp}/t = 0.25$.}
\label{fig:GN}
\end{figure}

We turn now to the nodal direction of the Brillouin zone. Figure~\ref{fig:GN} reveals
that  $G({\pmb k},0)$ remains almost unchanged with respect to the 1D regime up to the
${\pmb k}$-selective Mott transition at $t_{\perp}/t=0.18$. This agrees with: (i) the behavior of 
$\Sigma({\pmb k},i\omega_m\to 0)$ at ${\pmb k}=(\pi/2,\pi/2)$  which we expect to diverge in
the $T\to 0$ limit up to $t_{\perp}/t=0.2$,  and (ii) the gap seen in $A({\pmb k},\omega)$.
On further increasing $t_{\perp}$, $G({\pmb k},0)$ develops a pole-like feature. As a result,
a broad QP peak emerges at $t_{\perp}/t=0.25$. Its weight is much smaller than that at ${\pmb k}_F$ along the
$(0,0)-(\pi,0)$ direction included for comparison in Fig.~\ref{fig:GN}. The substantial variation in the 
scattering rate is a precursory indication of the broken FS at smaller $t_{\perp}$.

Let us expand briefly on a relationship of our results with real materials such as quasi-1D organic
(TM)$_2$X salts where TM stands for TMTTF (tetramethyltetrathiafulvalene) or
its selena TMTSF analog~\cite{Giamarchi04}. 
Their unified phase diagram bears similarities to that of the high-$T_c$ cuprates in the sense that as a function of 
pressure  a superconducting phase emerges in the proximity of the insulating Mott phase. 
The inclusion of a finite $t'$ in our studies mimics the unnesting role of pressure.  
As we  show, crossing over from the 1D system to a higher-dimensional regime involves recombination 
of spinons and holons into the conventional QPs and releases the charge from the confinement 
to the 1D chains.  
The pressure-induced change in the optical spectroscopy data on (TMTTF)$_2$X 
is interpreted as an example of such a  charge \emph{deconfinement}~\cite{Ves98, Pas10}.
A quantitative comparison would involve studies within a model Hamiltonian
extended by electron-phonon coupling and long-range Coulomb interaction to
account for a wide variety of phase transitions observed in the (TM)$_2$X
families. Nevertheless, we find encouraging that our CDMFT studies
within the bare Hubbard model Eq.~(\ref{eq:Hubb}) support accumulating
experimental evidence that superconductivity in the (TMTSF)$_2$X salts is mainly mediated by magnetic 
fluctuations~\cite{Jer12}. Finally, the signatures of the closed FS contours that
we find could be verified in the quantum oscillation experiments.

In summary, our CDMFT simulations  yield a continuous quantum phase  transition between a 1D Mott insulating state 
and a 2D metallic state.  On the metallic side the \emph{coherence} temperature below which QPs  form marks the 
crossover scale  and vanishes at the critical point. At energy scales  below
the  coherence temperature and in the close vicinity  of  the transition point,  
the FS topology  shows hole and electron pockets.  
We attribute their origin to the remnant 1D Umklapp  scattering at the nodal momenta.  
Such a mechanism can also account for the pockets observed  in the \emph{spinless} model~\cite{Bert06}.  
The evolution of the pockets with $t_{\perp}$ can be understood by tracking
the zero and poles of the single-particle Green's function.  
At energy scales  above the  coherence temperature, remnant features of spin-charge separation are apparent in the 
single-particle spectral function.  At the two-particle level, the  metallic
state is characterized by enhanced antiferromagnetic fluctuations in the very
close vicinity of the critical point.  These magnetic fluctuations act as a glue for paring 
correlations with a $d$-wave character.  On the insulating side, the crossover
scale is set by the Mott gap.  Below this energy scale we observe robust 1D Mott physics: 
aspects of spin-charge separation are visible both in the spectral function and $(\pi,0)$ magnetic 
fluctuations which remain intact.    
Further work  aimed at  investigating the finite
temperature consequences of this quantum critical point is presently under progress.

\begin{acknowledgments}
We acknowledge support from the DFG grant AS120/8-1 (Forschergruppe FOR 1346)  and thank the LRZ-M\"unich and 
the J\"ulich Supercomputing center for a generous allocation of CPU time.   
\end{acknowledgments}


\begin{thebibliography}{10}

\bibitem{Imada98}
M. Imada, A. Fujimori, and Y. Tokura, Rev. Mod. Phys. {\bf 70},  1039  (1998).

\bibitem{Lim03}
P. Limelette, A. Georges, D. J\'erome, P. Wzietek, P. Metcalf, and J.~M. Honig,
  Science {\bf 302},  89  (2003).

\bibitem{Kag05}
F. Kagawa, K. Miyagawa, and K. Kanoda, Nature {\bf 436},  534  (2005).

\bibitem{Kag09}
F. Kagawa, K. Miyagawa, and K. Kanoda, Nat. Phys. {\bf 5},  880  (2009).

\bibitem{Sent11}
M. Sentef, P. Werner, E. Gull, and A.~P. Kampf, Phys. Rev. B {\bf 84},  165133
  (2011).

\bibitem{Sem11}
P. S\'emon and A.-M.~S. Tremblay, Phys. Rev. B {\bf 85},  201101  (2012).

\bibitem{Sato11}
T. Sato, K. Hattori, and H. Tsunetsugu, e-print arXiv:1111.5371 (2011).

\bibitem{Sene04}
D. S\'en\'echal and A.-M.~S. Tremblay, Phys. Rev. Lett. {\bf 92},  126401
  (2004).

\bibitem{Civ05}
M. Civelli, M. Capone, S.~S. Kancharla, O. Parcollet, and G. Kotliar, Phys.
  Rev. Lett. {\bf 95},  106402  (2005).

\bibitem{Macr06}
A. Macridin, M. Jarrell, T. Maier, P.~R.~C. Kent, and E. D'Azevedo, Phys. Rev.
  Lett. {\bf 97},  036401  (2006).

\bibitem{Giamarchi_book}
T. Giamarchi, {\em Quantum Physics in One Dimension} (Oxford Univ. Press,
  Oxford, 2004).

\bibitem{Lake10}
B. Lake, A.~M. Tsvelik, S. Notbohm, D.~A. Tennant, T.~G. Perring, M. Reehuis,
  C. Sekar, G. Krabbes, and B. B\"uchner, Nat. Phys. {\bf 6},  50  (2010).

\bibitem{Geo96}
A. Georges, G. Kotliar, W. Krauth, and M.~J. Rozenberg, Rev. Mod. Phys. {\bf
  68},  13  (1996).

\bibitem{Kot01}
G. Kotliar, S.~Y. Savrasov, G. P\'alsson, and G. Biroli, 
Phys. Rev. Lett. {\bf 87}, 186401 (2001). 

\bibitem{Bol03}
C.~J. Bolech, S.~S. Kancharla, and G. Kotliar, Phys. Rev. B {\bf 67},  075110
  (2003).

\bibitem{Cap04}
M. Capone, M. Civelli, S.~S. Kancharla, C. Castellani, and G. Kotliar, Phys.
  Rev. B {\bf 69},  195105  (2004).

\bibitem{Bier01}
S. Biermann, A. Georges, A. Lichtenstein, and T. Giamarchi, Phys. Rev. Lett.
  {\bf 87},  276405  (2001).

\bibitem{Bert06}
C. Berthod, T. Giamarchi, S. Biermann, and A. Georges, Phys. Rev. Lett. {\bf
  97},  136401  (2006).

\bibitem{Tsu07}
M. Tsuchiizu, Y. Suzumura, and C. Bourbonnais, Phys. Rev. Lett. {\bf 99},
  126404  (2007).

\bibitem{Mou11}
S. Moukouri and E. Eidelstein, Phys. Rev. B {\bf 84},  193103  (2011).

\bibitem{sign}
The limiting factor comes from the $(\beta N_c)^3$ scaling of the QMC solver 
while for the considered parameters and cluster sizes the QMC sign problem remains 
very mild. 

\bibitem{Beach04a}
K.~S.~D. Beach, e-print arXiv:cond-mat/0403055 (2004).

\bibitem{Ess02}
F.~H.~L. Essler and A.~M. Tsvelik, Phys. Rev. B {\bf 65},  115117  (2002).

\bibitem{White89}
S.~R. White, D.~J. Scalapino, R.~L. Sugar, N.~E. Bickers, and R.~T. Scalettar,
  Phys. Rev. B {\bf 39},  839  (1989).

\bibitem{Mat05}
H. Matsueda, N. Bulut, T. Tohyama, and S. Maekawa, Phys. Rev. B {\bf 72},
  075136  (2005).

\bibitem{epaps1}
See Supplemental Material for the spectral function $A({\pmb k},\omega)$ 
on the $8\times 2$ cluster at $t_{\perp}/t=0.15$ and $t_{\perp}/t=0.2$.

\bibitem{Rib11}
P. Ribeiro, P.~D. Sacramento, and K. Penc, Phys. Rev. B {\bf 84},  045112
  (2011).

\bibitem{epaps2} 
See Supplemental Material for the evidence of the FS pockets on the $4\times 4$ cluster. 

\bibitem{Stan06}
T.~D. Stanescu and G. Kotliar, Phys. Rev. B {\bf 74},  125110  (2006).

\bibitem{Sak09}
S. Sakai, Y. Motome, and M. Imada, Phys. Rev. Lett. {\bf 102},  056404  (2009).

\bibitem{Gros12}
L.~F. Tocchio, F. Becca, and C. Gros,
Phys. Rev. B {\bf 86}, 035102 (2012).

\bibitem{Giamarchi04}
T. Giamarchi, Chem. Rev. {\bf 104},  5037  (2004).

\bibitem{Ves98}
V. Vescoli, L. Degiorgi, W. Henderson, G. Gr\"uner, K. Starkey, and L.~K.
  Montgomery, Science {\bf 281},  1181  (1998).

\bibitem{Pas10}
A. Pashkin, M. Dressel, M. Hanfland, and C.~A. Kuntscher, Phys. Rev. B {\bf
  81},  125109  (2010).

\bibitem{Jer12} 
D. J\'erome, J. Supercond. Nov. Magn. {\bf 25}, 633 (2012).

\end{thebibliography}

\section*{Supplementary data to the article: 
Dimensional-Crossover-Driven Mott Transition in the Frustrated Hubbard Model}

\begin{figure}[b!]
\begin{center}
\includegraphics*[width=0.45\textwidth]{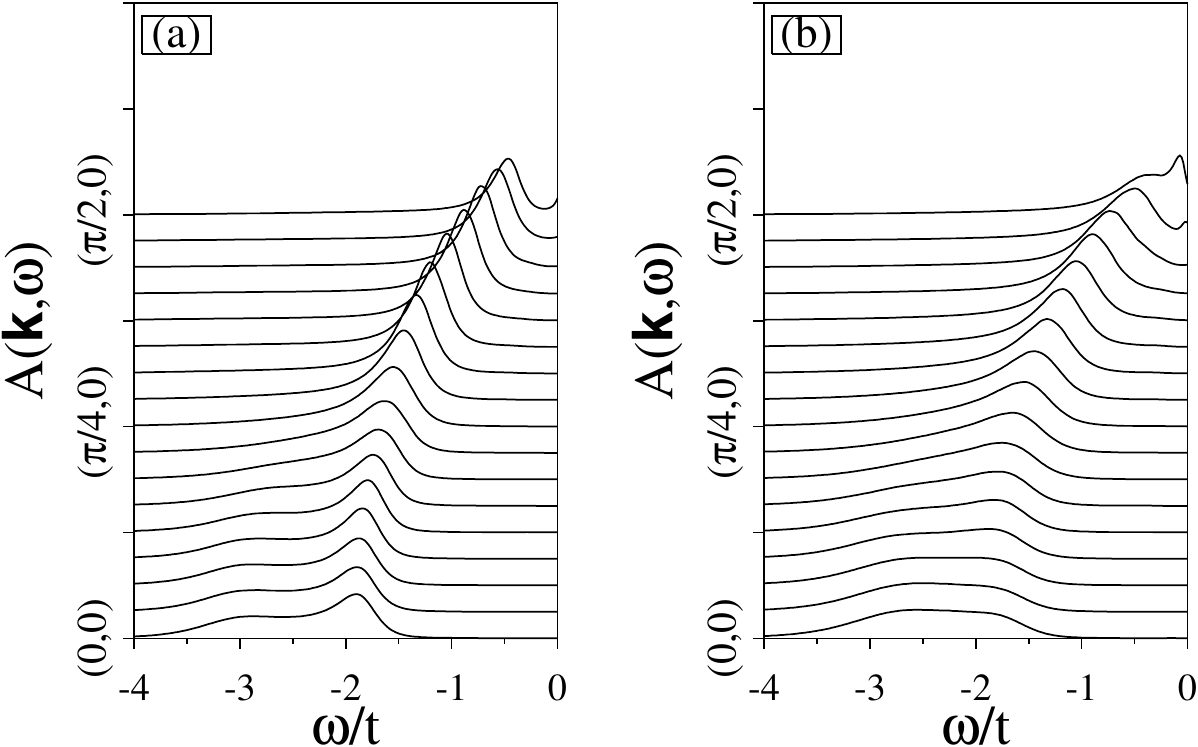}
\end{center}
\caption {
Dimensional-crossover-driven Mott transition as seen in the spectral function $A({\pmb k},\omega)$
within the CDMFT on the $8\times 2$ cluster: (a) $t_{\perp}/t=0.15$ and (b) $t_{\perp}/t=0.2$.
Parameters: $U/t=3$, $t'=-t_{\perp}/4$ and $T=t/30$.}
\label{fig:supl}
\end{figure}

Figure~\ref{fig:supl} shows the single-particle spectral function $A({\pmb k},\omega)$ on
the $8\times 2$ cluster for $t_{\perp}/t=0.15$ and $t_{\perp}/t=0.2$ and 
it complements Fig.~\ref{fig:Ak} in our Letter.   

\begin{figure}[b!]
\begin{center}
\includegraphics[width=0.45\textwidth]{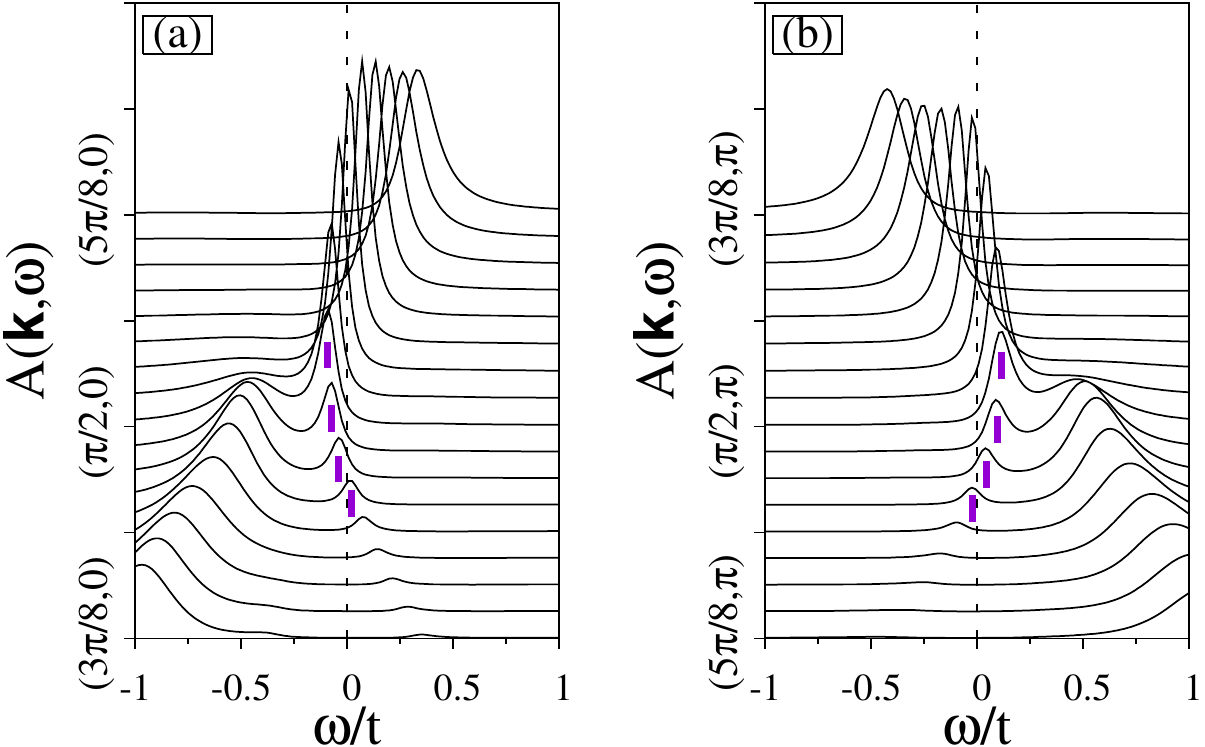}
\end{center}
\caption {(color online)
Low-energy part of the spectral function $A({\pmb k},\omega)$ at
$t_{\perp}/t=0.2$ within the CDMFT on the $4\times 4$ cluster.
Vertical bars track the position of the peaks constituting:
(a) electron and (b) hole FS pockets. Parameters: $U/t=3$, $t'=-t_{\perp}/4$ and $T=t/30$. }
\label{fig:supl2}
\end{figure}

Auxiliary CDMFT simulations performed on the $4\times 4$ cluster indicate that the
broken FS is unrelated to a specific ladder geometry of the $8\times 2$ cluster (c.f. Fig.~\ref{fig:supl2}) 
Hence, FS pockets appear to be a generic low-energy feature
accompanying the Mott transition in the quasi-1D limit.

\end{document}